\documentstyle[pra,preprint,aps,eqsecnum,tighten]{revtex}

\def\bra#1{\langle #1 |}
\def\ket#1{| #1 \rangle}
\def\bracket#1#2{\langle #1 | #2 \rangle}
\def\expect#1{\langle #1 \rangle}
\def\ident{{\hat 1}}
\def\H{{\hat H}}
\def\Heff{{\H_{\rm eff}}}
\def\P{{\hat {\cal P}}}
\def\C{{\hat C}}
\def\R{{\hat R}}
\def\F{{\hat F}}
\def\L{{\cal L}}

\def\Lhat{{\hat L}}
\def\e{{\rm e}}
\def\Tr{{\rm Tr}}
\def\a{{\hat a}}
\def\adag{{\hat a}^\dagger}
\def\b{{\hat b}}
\def\bdag{{\hat b}^\dagger}
\def\c{{\hat c}}
\def\cdag{{\hat c}^\dagger}
\def\r{{\hat r}}
\def\rdag{{\hat r}^\dagger}
\def\n{{\hat n}}
\def\d{^\dagger}
\def\rhot{{\tilde\rho}}
\def\Dbar{{\bar D}}

\begin{document}

\title{Continuous measurements, quantum trajectories, \\
and decoherent histories}

\author{Todd A. Brun\cite{Email} \\
Department of Physics, Carnegie Mellon University, \\
Pittsburgh, PA 15213}

\date{\today}

\draft

\preprint{NSF-ITP-97-116}

\maketitle

\begin{abstract}
Quantum open systems are described in the Markovian limit by master
equations in Lindblad form.  I argue that common ``quantum trajectory''
techniques corresponding to continuous measurement schemes, which solve
the master equation by unraveling its evolution into stochastic
trajectories in Hilbert space, correspond closely to particular sets
of decoherent (or consistent) histories.
This is illustrated by a simple model of photon counting.
An equivalence is shown for these models
between standard quantum jumps and the orthogonal jumps of Di\'osi,
which have already been shown to correspond to decoherent histories.
This correspondence is compared to simple
treatments of trajectories based on repeated or continuous measurements.
\end{abstract}

\pacs{03.65.Bz 03.65.Ca}

\section{Introduction}

Recently a great deal of work has been done in
quantum optics on simulations of continuously
measured systems with dissipation, referred to variously as 
quantum jumps, relative state, and Monte Carlo Wavefunction
simulations \cite{Carmichael1,Dalibard,Gardiner,Wiseman1,Carmichael2,Plenio},
which are examples of a class of techniques known as
``quantum trajectories.''  In these techniques, a deterministic
master equation for the density matrix of an open system is replaced
with a stochastic differential equation for a pure quantum state.
Averaging the solutions over all realizations of the noise reproduces
the master equation.  Such a stochastic pure-state equation
is known as an ``unraveling'' of the master equation.  Unraveling is
not unique; in general, there can be many stochastic equations which
average to the same master equation.  A single solution for one realization
of the noise is called a ``quantum trajectory.''

In the original conception, the system was assumed to be monitored by
continuous measurements performed on the environment.  The information
from the measurements ``collapsed'' the system density matrix to
a pure state, but the randomness of the measurement outcomes made this
state unpredictable.  The effective description of the continuously
monitored system is thus a stochastic differential equation:  an unraveling.
Different unravelings correspond to different measurement schemes.
Averaging over all possible measurement outcomes reproduces the master 
equation, which is the best prediction that could be made in the absence
of any measurements.

However, even with no knowledge of the environment and its interaction
with the system, one can formally unravel any master equation into
a stochastic differential equation for pure states.  In this case,
the unraveling is simply used as a means of solving the master equation.
This provides a useful numerical technique, commonly called ``quantum
Monte Carlo.''  By averaging over many solutions of the quantum
trajectory one can find an approximate solution to the master equation.
A density operator $\rho$ on a Hilbert space of dimension $N$ requires
$N^2-1$ real numbers to represent it; this can be computationally
prohibitive for large $N$, while a single state (of size $2N-2$)
remains practical, even with the necessity
of averaging over many runs of the stochastic equation.

Around the same time that quantum trajectories were introduced,
the consistent (or decoherent) histories formulation of quantum mechanics
was developed by Griffiths, Omn\`es, and Gell-Mann and Hartle
\cite{Griffiths,Omnes1,Omnes2,GMHart1,GMHart2,Omnes3}.
In this formalism, one describes a quantum system in terms of an exhaustive
set of possible histories, which must satisfy a {\it decoherence}
or {\it consistency} criterion.
Histories which satisfy this criterion do not interfere with each other,
and may be assigned probabilities which obey the usual classical
probability sum rules.

Both quantum trajectories and consistent histories describe a quantum system
in terms of alternative possible evolutions; they thus bear a certain
resemblance to each other.  What is more, sets of histories corresponding
to possible records of a ``classical'' measuring device will
always decohere.  Thus, there will be a set of
consistent histories which correspond to the quantum trajectories of a
continuously measured system.

Exactly such a correspondence has been shown between decoherent
histories and quantum state diffusion (QSD), a particular unraveling of
the master equation, by Di\'osi, Gisin, Halliwell and Percival \cite{DGHP}.
QSD trajectories were shown to correspond to a set of approximately
consistent histories for a specific choice of projection operators
at closely spaced intervals of time.  Earlier, Di\'osi \cite{Diosi1} had
shown that a particular type of branch-dependent decoherent history
first examined by Paz and Zurek \cite{PazZurek} corresponded to yet another
type of trajectory, the orthogonal jump unraveling.
Below I will show a
similar correspondence for quantum jumps, and I conjecture that
most useful unravelings correspond to some set of decoherent histories
in an analogous way.  I also demonstrate an interesting correspondence
between standard quantum jumps and the orthogonal jumps of Di\'osi.
I have shown the correspondence with decoherent histories
for a simple model in an earlier
paper \cite{Brun}, and there has also been work by Ting Yu
\cite{Yu} from a rather different perspective on the relationship of
quantum jumps and decoherent histories.  In this paper, I will analyze
the correspondence between quantum trajectories and decoherent histories
in general using the concept of {it generalized records} introduced
by Gell-Mann and Hartle \cite{GMHart3}, which provides a unifying
framework for both.  Recent work on non-Markovian quantum state
diffusion explicitly includes the correlation between the trajectory
and the state of the environment \cite{Diosi3,Strunz}; in that case,
the environment state is precisely this kind of generalized record.

In section II, I review the formalism of quantum trajectories, giving
the quantum jump unraveling in some detail but also briefly describing
the quantum state diffusion and ortho-jump unravelings.  I then
review the formalism of decoherent (consistent) histories, and
contrast it with quantum trajectories.  I point out that the usual
quantum trajectory description itself assumes a certain consistency
condition.

In section III, I examine a simple model of a photon counting experiment,
in which the photodetector is represented by a Markovian
environment producing rapid dissipation and decoherence in a field
mode coupled to a quantum system.  This representation reproduces
the effects of repeated Von Neumann measurements, and in
the limit of separated timescales between the system and environment
allows one to derive a Markovian
master equation for the system alone.  This master equation can be
unraveled into quantum jump or ortho-jump
trajectories as described in sections IA and IB above, and we will
see that there is an equivalence between these unravelings for this model.

In section IV, I start with the same model of a photon counting experiment,
but this time enumerate a set of consistent histories representing
different photodetection records.  We see that the probabilities of
these histories are the same as the quantum jump trajectories in section
III, and that these histories do decohere to a high level of accuracy.

In section V, I show that the model detector of section III produces
generalized records, the existence of which guarantees that the
consistency conditions will be met.  The existence of generalized records
allows the interpretation of quantum trajectories as consistent histories
even in a case with no measurement devices; I argue this in the case
when the outgoing electromagnetic field from the system itself serves
as a generalized record (though one with many incompatible interpretations).

Finally, in section VI, the results are summarized, and we see that they both
generalize the notion of quantum trajectories and provide a useful
calculational tool for the formalism of decoherent histories.

\section{Quantum trajectories and decoherent histories}

\subsection{Quantum Trajectories}

The systems of interest in quantum trajectories are
described by a Lindblad master equation in the Markov
approximation \cite{Lindblad},
\begin{equation}
{\dot \rho} = - i [\H,\rho] + \sum_m \left( \Lhat_m \rho \Lhat_m^\dagger
  - {1\over2} \Lhat_m^\dagger \Lhat_m \rho
  - {1\over2} \rho \Lhat_m^\dagger \Lhat_m \right)\ ,
\label{master_eqn}
\end{equation}
where $\rho$ is the reduced density operator
of the system, $\H$ is the system Hamiltonian, and the $\{\Lhat_m\}$ are a
set of {\it Lindblad operators} which model the effects of the environment.
(Note that throughout this paper I use units where $\hbar=1$.)
One of the notable characteristics of master equations
of the form (\ref{master_eqn}) is that
they do not, in general, preserve pure states.  Suppose that the system
is initially in the state
\[
\rho(0) = \ket{\psi_0} \bra{\psi_0}.
\]
Over time it will evolve into a mixture which can be written
\begin{equation}
\rho(t) = \sum_\xi \ket{\psi_\xi(t)} p_\xi \bra{\psi_\xi(t)}
  \equiv {\rm M}(\ket{\psi_\xi(t)}\bra{\psi_\xi(t)}),
\label{rho_expansion}
\end{equation}
where ${\rm M}()$ is the ensemble mean with probabilities
\[
p_\xi \ge 0,\ \ \sum_\xi p_\xi = 1.
\]
The decomposition (\ref{rho_expansion}) is generally not unique; there
can be many different sets of states $\{\ket{\psi_\xi}\}$ which give the
same density operator $\rho$.  This ambiguity leads to an ambiguity in
unraveling the evolution, to which we will return shortly.

Now let us choose a set of states
$\ket{\psi(t,\xi(t))}$, where $\xi(t)$ is a random process and
$\ket{\psi(t,\xi(t))}$ obeys a stochastic evolution equation.
By averaging over all solutions of this stochastic equation
for all realizations of the noise $\xi(t)$, one finds the solution
of (\ref{rho_expansion}) at time $t$.
This stochastic equation will be of the form
\begin{equation}
\ket{d\psi} = \ket{u} dt + \ket{v} d\xi(t),
\end{equation}
where $d\xi(t)$ is a stochastic differential variable representing
the random process $\xi(t)$, and can include continuous diffusive noise,
discrete jumps, or both.  The vectors $\ket{u}$ and $\ket{v}$
are functions of the state $\ket\psi$.  (In general, there will be
several noise terms with different $d\xi(t)$.)
A single solution $\ket{\psi(t,\xi(t))}$ of this equation
for a single realization of $\xi(t)$
follows a quantum trajectory in Hilbert space.
The complete set of solutions for all $\xi(t)$
constitutes an unraveling of the master equation.

This is a rather formal notion of quantum trajectories, since it is not
clear what physical significance the states $\ket{\psi(t,\xi(t))}$
have, if any.  Given the existence of many different expansions
(\ref{rho_expansion}) for the same density operator $\rho(t)$, one must
also admit the possibility of many different possible unravelings.  If
the choice of unraveling is not unique, what
physical reality does it have?  Fortunately, it is
possible to use the abstract idea of unraveling in practical
situations where  $\ket{\psi(t,\xi(t))}$ {\it does} have a physical
interpretation.  Indeed, this was the original motivation for developing
quantum trajectories \cite{Carmichael2}.

The non-unitarity of the master equation is caused by the loss
of information from the system to the environment, as their interaction
produces entanglement between the system and
environment degrees of freedom.  Within the master equation
description this entanglement is inaccessible;
but if one has experimental access to the environment, it is possible to
extract information about the system by making selected
measurements.  If the measurements are performed with sufficient precision
and frequency, and the initial state is known, one can determine
the exact state of the system at later times.

If we include the environment in our description, the joint
system-environment state remains pure:
\begin{equation}
\ket\Psi = \sum_i c_i \ket{a_i} \otimes \ket{b_i},
\label{purification}
\end{equation}
where $\ket{a_i}$ and $\ket{b_i}$ are states in the system and environment
Hilbert spaces, respectively.  For any state $\ket\Psi$ it is possible
to find an expression (\ref{purification}) such that the $\ket{a_i}$
are all mutually orthogonal, as are the $\ket{b_i}$; this is the
Schmidt decomposition \cite{Peres}.
However, in general this is not necessary; any
decomposition (\ref{purification}) in which the $\ket{b_i}$  (but not
necessarily the $\ket{a_i}$) are orthogonal will suffice.  In particular,
given an observable for the environment, we can choose the $\ket{b_i}$
to be a basis of eigenstates.  Then by measuring the environment, the
system goes into the state $\ket{a_i}$ with probability $|c_i|^2$.

In this picture the system remains in a pure state, but its evolution is
no longer deterministic:  it is influenced by random measurement
outcomes.  This is often referred to as {\it conditional}
(or {\it relative state}) evolution:  it is conditioned on the measurement
record.  If one averages over all possible measurement outcomes,
a particular decomposition of the
density operator (\ref{rho_expansion}) results.
This exactly matches the earlier notion of a quantum trajectory,
but now the unraveling has a clear-cut physical interpretation:
the state $\ket{\psi(t,\xi(t))}$ represents our knowledge of the system,
conditioned on the random outcomes $\xi(t)$ of a sequence of measurements,
and $p_\xi$ is the probability of this outcome.

Note that not all measurements will satisfy the requirement
(\ref{rho_expansion})!  Indeed, only a fairly restricted subclass will
do so.  We discuss this restriction in IIE below, and later in section V.

\subsection{Quantum Jumps}

We can make this clearer with a concrete example.  Suppose
we consider a quantum optical system, such as a small cavity with
one mirror partially silvered so that radiation can
escape.  The external electromagnetic field represents the environment 
in this case, while a field mode inside the cavity is the
system.  We describe the system evolution by a master equation
\cite{Carmichael2}
\begin{equation}
{\dot \rho} = - i [\H_0,\rho] + \gamma \a \rho \adag
  - (\gamma/2) \adag \a \rho
  - (\gamma/2) \rho \adag \a,
\label{cavity}
\end{equation}
where $\H_0$ represents the Hamiltonian of the system mode, $\a$ is the
annihilation operator, and $\gamma$ represents the rate of dissipation from
the cavity.  We will be considering this simple master equation
for most of this paper.

Equation (\ref{cavity}) describes the evolution of the system {\it provided}
that we know nothing about the quanta which escape to the environment,
or choose to ignore that information.
Suppose we place photodetectors outside the cavity in such a way
that every escaped photon is detected.  Each detection gives us some
information about the state of the system.  In this case, with perfect
detection, the system evolution becomes
\begin{equation}
{d\ket\psi\over dt} = - i \H_{\rm eff} \ket\psi,
\label{qj_schrod}
\end{equation}
interrupted at random times by sudden quantum jumps
\begin{equation}
\ket\psi \rightarrow \a\ket\psi,
\label{jump}
\end{equation}
where $\H_{\rm eff}$ is the {\it effective Hamiltonian}
\begin{equation}
\H_{\rm eff} = \H_0 - i (\gamma/2) \adag\a,
\end{equation}
and the jumps occur with a probability given by the norm of the state,
as we shall see explicitly below \cite{Carmichael1,Dalibard,Gardiner}.
Equation (\ref{qj_schrod}) has the form of the usual
Schr\"odinger equation, but with a non-hermitian Hamiltonian.
A state that evolves for a time $t$ without jumping is given by
\begin{equation}
\ket{\psi(t)} = \e^{-i\H_{\rm eff} t} \ket{\psi(0)}.
\end{equation}

The jumps represent detections of emitted photons, which trigger
a sudden change in our knowledge of the system.  The presence of the
non-Hermitian terms in the effective Hamiltonian represents the effect
of {\it not} detecting any photons on our knowledge of the system; a
null measurement thus still affects the system.

This evolution does not preserve the norm of the state.
The actual physical state is
taken to be $\ket{\tilde\psi} = \ket\psi/\sqrt{\bracket{\psi}{\psi}}$,
the renormalized state.  An actual physical detector cannot determine
the time of a photon emission with infinite precision; at best, it will
determine the time within a short interval $\Delta t$.
The probability that an initial state $\ket\psi$ evolves for a time $T$
and undergoes $N$ jumps at times $t_1, \ldots, t_N$
(which are assumed to be widely spaced with respect to
$\Delta t$) is
\begin{eqnarray}
p(\ket{\tilde\psi}) = &  (\gamma\Delta t)^N
  \Tr\biggl\{ \e^{-i\Heff(T-t_N)} \a \e^{-i\Heff(t_N - t_{N-1})} \a
  \cdots \a \e^{-i\Heff t_1} \nonumber\\
& \times \ket\psi\bra\psi \e^{i\Heff^\dagger t_1} \adag
  \cdots \adag \e^{i\Heff^\dagger(T-t_N)} \biggr\}.
\label{jumps_prob}
\end{eqnarray}
This trace is the norm of the state
\begin{equation}
\ket{\psi_{t_1,\ldots,t_N}} =
  \e^{-i\Heff(T-t_N)} \a \e^{-i\Heff(t_N - t_{N-1})} \a
  \cdots \a \e^{-i\Heff t_1} \ket\psi,
\end{equation}
which is the unrenormalized state resulting from
this particular sequence of jumps $t_1, \ldots, t_N$.  The norm of the
unrenormalized state thus gives the probability density for that
state to be realized.  The density operator $\rho$ is given by averaging
over all realizations $\ket{\tilde\psi}$ with probability
measure (\ref{jumps_prob}),
\begin{eqnarray}
\rho(t) = && {\rm M}(\ket{\tilde\psi(t)} \bra{\tilde\psi(t)}) \nonumber\\
= && \sum_{\ket{\tilde\psi}} \ket{\tilde\psi} p(\ket{\tilde\psi})
  \bra{\tilde\psi} \nonumber\\
= && \sum_{\ket{\psi_{t_1,\ldots,t_N}}} (\gamma\Delta t)^N
  \ket{\psi_{t_1,\ldots,t_N}} \bra{\psi_{t_1,\ldots,t_N}}, \nonumber\\
= && \sum_N \gamma^N \int_0^T dt_1 \cdots \int_{t_{N-1}}^T dt_N
  \ket{\psi_{t_1,\ldots,t_N}} \bra{\psi_{t_1,\ldots,t_N}}.
\label{mean}
\end{eqnarray}
The density operator defined by (\ref{mean}) solves the master equation
(\ref{cavity}).

It is possible to rewrite these equations in
explicitly norm-preserving form,
\begin{equation}
\ket{d\tilde\psi} = - i \H_{\rm eff} \ket{\tilde\psi} dt
  + (\gamma/2)\expect{\adag\a} \ket{\tilde\psi} dt
  + \left( {\a\over{\sqrt{\expect{\adag\a}}}} - 1 \right) \ket{\tilde\psi} dN,
\label{nonlinear_jumps}
\end{equation}
at the cost of a little extra complexity
and nonlinearity \cite{Dalibard}.  This is a stochastic differential
equation, where $dN$ is a stochastic variable which is 0 except at
random times (corresponding to the jumps) when it becomes 1.  It
has statistics
\begin{equation}
dN dN = dN,\ \ {\rm M}_{\ket\psi}(dN) = \gamma\expect{\adag\a},
\end{equation}
where ${\rm M}_{\ket\psi}$ denotes the ensemble average over all trajectories
which are in state $\ket\psi$ at the given time $t$.
The nonlinear form has useful properties, but
for our purposes the linear form will usually prove more convenient.

\subsection{Other Unravelings}

The quantum jump equation is not the only way of
unraveling the master equation into
quantum trajectories.  In fact, there are an infinite number
of such unravelings.  Certain choices are more commonly used than others,
however, so we will go over them quickly.

One of the most useful is the {\it quantum state diffusion}
(QSD) equation \cite{GisinPercival}
\begin{equation}
\ket{d\psi} = - i \H \ket\psi dt
  + \sum_m \left( \expect{\Lhat^\dagger_m} \Lhat_m
  - {1\over2} \Lhat^\dagger_m \Lhat_m
  - {1\over2} | \expect{\Lhat_m} |^2 \right) \ket\psi dt
  + \sum_m \left( \Lhat_m - \expect{\Lhat_m} \right) \ket\psi d\xi_m.
\label{qsd_eqn}
\end{equation}
This is an It\^o stochastic differential equation \cite{Gardiner2},
with the $d\xi_m$ representing continuous complex
stochastic processes with ensemble means
\begin{equation}
{\rm M}(d\xi_m) = {\rm M}(d\xi_m d\xi_n) = 0,\ \ 
  {\rm M}(d\xi_m d\xi^*_n) = \delta_{mn} dt.
\end{equation}
It is not difficult to show that this equation also obeys the master
equation (\ref{master_eqn}) in the mean:
\begin{equation}
\rho = {\rm M}(\ket{\psi_\xi}\bra{\psi_\xi}).
\end{equation}
Rather than discrete jumps, the solutions of
this equation undergo continuous diffusion.

This standard form of the QSD equation is nonlinear and norm-preserving.
However, there is a different unraveling known as
{\it linear QSD} with equation
\begin{equation}
\ket{d\psi} = - i \H \ket\psi dt
  - {1\over2} \sum_m \Lhat^\dagger_m \Lhat_m \ket\psi dt
  + \sum_m \Lhat_m \ket\psi d\xi_m.
\label{linear_qsd}
\end{equation}
This does {\it not} preserve the norm of $\ket\psi$, but is also
a valid unraveling of the master equation, and has properties similar
to the nonlinear equation \cite{linearQSD}.
 
The QSD equation was discovered by Gisin and Percival, following from
earlier work by Gisin in the theory of measurement.  
Carmichael \cite{Carmichael2} showed that a similar diffusion equation 
arose in the case (\ref{cavity}) from a relative state approach, just 
as in quantum jumps, but with direct photodetection replaced by
balanced homodyne detection.  It was shown by Wiseman
and Milburn \cite{Wiseman2} and others \cite{Knight} that the exact
QSD equation arises in the case of balanced heterodyne detection.
Thus, one can consider both approaches as giving the state
of a system conditioned on a measurement record, but using different
measurement schemes.  We will discuss this further below.  (The
linear equation (\ref{linear_qsd}) does not have such a straightforward
interpretation in terms of measurements, though that does not necessarily
imply that such an interpretation might not be found.)

Conversely, Gisin and Percival have shown \cite{GisinPercival}
that jump-like behavior can
be exhibited by the QSD equation by explicitly including a portion of
the photodetector in addition to the system; their explicit inclusion
of part of the environment is similar to the approach of this paper.

Related to QSD are the {\it orthogonal jumps} of Di\'osi \cite{Diosi2}.
These trajectories obey an equation
\begin{eqnarray}
\ket{d\psi} = && - i \H \ket\psi dt
  + \sum_m \left( \expect{\Lhat^\dagger_m} \Lhat_m
  - {1\over2} \Lhat^\dagger_m \Lhat_m
  + {1\over2} \expect{\Lhat_m^\dagger \Lhat_m}
  - | \expect{\Lhat_m} |^2 \right) \ket\psi dt \nonumber\\
&& + \sum_m \left( \ket{\psi_m} - \ket\psi \right) dN_m,
\label{ortho_jumps}
\end{eqnarray}
where the $dN_m$ now represent a stochastic jump process,
\begin{equation}
dN_m dN_n = \delta_{mn} dN_m,\ \ {\rm M}(dN_m) = r_m dt
\end{equation}
and the set of states $\ket{\psi_m}$ are mutually orthogonal and 
orthogonal to $\ket\psi$.  The formula for these orthogonal states
and their jump rates $r_m$ is complicated; fortunately, for the purposes
of this paper we will not need to worry about it.  The deterministic
portion of equation (\ref{ortho_jumps}) is identical to that of the
QSD equation expressed in Stratonovich rather than It\^o form
\cite{Rigo}.

These orthogonal jumps are in some ways the most economical unraveling
of the master equation, in that these jumps occur fairly infrequently
but cause a large change in the state.
This unraveling also has close ties to decoherent histories, as we
shall see.  However, relatively little work has been
done on the connection between ortho-jumps and measurements
\cite{Breslin1,Breslin2}.
An important result is that ortho-jumps result from choosing the
Schmidt decomposition of (\ref{purification}), where the $\ket{a_i}$
as well as the $\ket{b_i}$ are all orthogonal, and measuring the
eigenbasis $\ket{b_i}$.  This measurement gives the maximum average
information about the system, but requires a dynamic measurement
scheme, since the basis $\ket{b_i}$ will in general be different
at different times.

This form of ortho-jumps is also explicitly norm-preserving, which
results in a nonlinear stochastic equation, just as for QSD.
Unlike the other equations presented, there is no completely linear
version of this unraveling; the orthogonalization procedure used in
determining the $\ket{\psi_m}$ is intrinsically nonlinear.  However,
for certain special problems these $\ket{\psi_m}$ are linear functions
of $\ket\psi$.  In this special case it is possible to find a linear
version of ortho-jumps which does not preserve the norm, and which
is closely related to the standard quantum jump equation.  We will
examine such a case below.

All of these unravelings correspond to the same Markovian master
equation.  If considered simply as a way of solving for the density
operator, it doesn't matter which unraveling one picks; averaging over
any set of trajectories will give the same final result.  But
if one actually has a physical description of the environment, and
of the measurements being performed on it, this equivalence
is broken, and a particular choice of unraveling is singled out.  This
is the case in this paper, where we assume a particular (albeit simplified)
description of the environment, including the effects of a measurement
apparatus, to model photodetection.

\subsection{Decoherent (or Consistent) Histories}

Let us now turn to consistent histories.
In nonrelativistic quantum mechanics one can specify a set of
possible histories by choosing a sequence of times $t_1, t_2, \ldots, t_n$
and at each time $t_i$ a complete set of orthogonal projection operators
$\{\P^j_{\alpha_j}(t_j)\}$, such that
\begin{equation}
\P^j_{\alpha_j}(t_j) \P^j_{\alpha'_j}(t_j)
  = \P^j_{\alpha_j}(t_j) \delta_{\alpha_j \alpha'_j},\ \ 
\sum_{\alpha_j} \P^j_{\alpha_j}(t_j) = \ident,
\end{equation}
where the $\P$'s are Heisenberg operators.
These projectors represent a complete set of exclusive alternatives at
the given times.  A single history corresponds to a choice of
one projection operator $\P^i_{\alpha_i}(t_i)$ at each time $t_i$.  This
history can be represented by the sequence of indices $\alpha_1, \ldots,
\alpha_n$, which I will denote by $h$.

This is not the most general type of history.  For instance, one
can make the set of projections at a later time depend on the choices
$\{\alpha_i\}$ at earlier times, making the histories
{\it branch-dependent}.  We will not need this extra generality in this
paper, but it can be useful in describing measurements which are
conditioned on the results of earlier measurements.

The decoherence (or consistency) criterion is
described by the {\it decoherence functional}
$D[h,h']$, a complex functional on pairs of histories.
Two histories $h$ and $h'$
are said to {\it decohere} if they satisfy the relation
\begin{equation}
D[h,h'] = p(h) \delta_{hh'},
\label{decoherence}
\end{equation}
where $p(h)$ defines to be the probability of history $h$.
A set of histories $\{h\}$
is said to be complete and consistent if all pairs of histories satisfy
(\ref{decoherence}) and their probabilities sum to unity.

We define a {\it history operator} $\C_h$
\begin{equation}
\C_h = \P^n_{\alpha_n}(t_n) \P^{n-1}_{\alpha_{n-1}}(t_{n-1}) \cdots
  \P^1_{\alpha_1}(t_1).
\label{history_op}
\end{equation}
In terms of this operator the decoherence functional becomes
\begin{equation}
D[h,h'] = \Tr \{ \C_h \rho_0 \C_{h'}\d \},
\end{equation}
where $\rho_0$ is the initial state.  If $\rho_0$ is pure,
$\rho_0 = \ket\psi\bra\psi$, then $D[h,h']$ is an inner product
\begin{equation}
D[h,h'] = \bra\psi\C_{h'}\d \C_h\ket\psi,
\label{inner_product}
\end{equation}
and the consistency condition (\ref{decoherence})
amounts to the assertion that the states
$\{\C_h\ket\psi\}$ are orthogonal for different $h$, with the
probability of a history given by the norm of the corresponding state.

\subsection{Consistency of quantum trajectories}

It is often convenient to consider these projections in the Schr\"odinger
rather than the Heisenberg picture.  In this case, the $\P$'s are no longer
time-dependent, and instead the time-evolution is given explicitly:
\begin{equation}
\C_h = \P^n_{\alpha_n} \e^{-i\H(t_n - t_{n-1})} \P^{n-1}_{\alpha_{n-1}}
  \e^{-i\H(t_{n-1} - t_{n-2})} \cdots \P^1_{\alpha_1} \e^{-i\H(t_1 - t_0)}.
\label{schrodinger}
\end{equation}
Written in this form, it is clear that the state $\C_h\ket\psi$ has a
strong resemblance to our previous definition of a quantum trajectory.
This succession of Hamiltonian and projection terms resembles some kind
of pure state evolution, and if no single history
has probability 1 this evolution will have a stochastic component.

However, there are a number of important differences.  The first
is the most obvious:  in our definition of quantum trajectories, there was
no requirement that they should obey the decoherence criterion.  This is
related to the fact that the decomposition (\ref{rho_expansion}) need not
be in terms of orthogonal states.  More on this below.

The second difference is that quantum trajectories were
framed in terms of a split between system and environment degrees
of freedom, with the environment traced out of the equations of motion
in the Markovian approximation.  These assumptions are not made in
consistent histories, which can be quite general.

The open systems case is important, however
\cite{FeynVern,Zurek,JoosZeh,CaldLegg}, and is worth treating in depth
in consistent histories.  Most such treatments to date 
\cite{GMHart2,Brun3,DowkHall} involve choosing a set of projection
operators on the system alone, while making
no assertions about the environment degrees of
freedom.  That is, one specializes to projections of the form
\begin{equation}
\P = \P_{\rm sys} \otimes \ident_{\rm env},
\label{sys_projector}
\end{equation}
where $\P_{\rm sys}$ is a projection operator in the Hilbert space of
the system and $\ident_{\rm env}$ is the identity operator in the Hilbert
space of the environment.  The Markovian approximation will only be
valid for certain choices of environments and of the system/environment
interaction, of course; but if one is treating the same systems in both
the quantum trajectories and decoherent histories approach, the approximation
should be good in both cases.  One ends up with a reduced description
in terms of the system alone, quite in the spirit of the master equation.

If we physically interpret the quantum trajectories as evolution
conditioned on measurements of the environment, we realize that
(\ref{sys_projector}) is not adequate to describe this situation.  To
make statements about the state of the environment, we require
projectors onto the environment, not just the system.  In fact, the
correct projectors in this case have the form
\begin{equation}
\P = \ident_{\rm sys} \otimes \P_{\rm env},
\label{env_projector}
\end{equation}
quite different from the usual discussion of decoherence in open systems.
Thus, to look for a set of histories equivalent to a particular
unraveling, we must retain in our description enough of the environment
to include suitable projections of the form (\ref{env_projector}).
As we shall see, in some cases this may be very little.  To reduce
to a description of the system alone, a further tracing out of the
environment is then required.

As far as consistency is concerned, it has not been widely appreciated
that quantum trajectories (as they are commonly defined) must
obey a consistency condition of their own.  This is easily seen 
by considering the effects of environmental measurements.  Rather
than tracing out the environment to get a master equation (\ref{master_eqn}),
we retain a complete description of the system and environment
(\ref{purification}).  There is a Hamiltonian $\H$ for the system and
environment together, such that they tend to become entangled with time:
\begin{equation}
\ket{\Psi(t)} = \e^{-i\H t} \ket{\Psi(0)}.
\end{equation}
Suppose we now perform a series of $n$ measurements on the environment
spaced $\Delta t = t/n$ apart in time,
with outcomes corresponding to projections of the form (\ref{env_projector}).
The (unnormalized) joint state conditioned on these outcomes then becomes
\begin{equation}
\ket{\Psi_\xi} = \P_{\xi_n} \e^{-i\H \Delta t} \P_{\xi_{n-1}}
  \cdots \P_{\xi_1} \ket{\Psi(0)},
\end{equation}
with probability $p_\xi = \bracket{\Psi_\xi}{\Psi_\xi}$.  We now want
to impose the condition (\ref{rho_expansion}).  In order for this
to hold we must have
\begin{equation}
\sum_\xi \Tr_{\rm env} \left\{ \ket{\Psi_\xi}\bra{\Psi_\xi} \right\} =
  \sum_{\xi,\xi'} \Tr_{\rm env} \left\{
  \ket{\Psi_\xi}\bra{\Psi_\xi'} \right\},
\end{equation}
which implies
\begin{equation}
  \sum_{\xi \ne \xi'} \Tr_{\rm env} \left\{
  \ket{\Psi_\xi}\bra{\Psi_\xi'} \right\} = 0.
\label{traj_consistency}
\end{equation}
This equation is by no means automatically satisfied.  It is, in fact,
a consistency condition rather like (\ref{decoherence}).  In
some ways it is much weaker, since (\ref{traj_consistency}) is only
imposed on sums of off-diagonal terms rather than individual terms
(although it must be satisfied for all times $t$).  In another way
it is stronger:  the trace in (\ref{traj_consistency}) is only over
the environment, so that an entire operator on the system space is
required to vanish, not just the trace.  This is rather like
the partial trace decoherence of Finkelstein \cite{Finkelstein},
discussed in section IV.  In physical terms, (\ref{traj_consistency})
means that the measurements made must not alter the state of the
environment such that it then acts back and alters the dynamics
of the system.  As we shall see, quantum jumps (like most of the commonly
considered unravelings) satisfies both (\ref{traj_consistency}) and
(\ref{decoherence}), and in fact satisfies a consistency criterion
stronger than either.  The Born-Markov approximation which is generally
made in deriving the master equation explicitly assumes that the state
of the environment is not changed by interacting with the system; so
long as measurements are made only on the outgoing field,
(\ref{traj_consistency}) will be satisfied.

What makes a set of histories decohere?  While a general characterization
is still the subject of research, common mechanisms of decoherence 
have been widely studied in simple models, and are
now fairly well understood.  Suppose that the state of the system
at the time $t_i$ becomes correlated with some degrees of freedom of
the environment described by a Hilbert subspace ${\cal H}_i$, and the
projection operator $\P_{\alpha_i}^i$ singles out a particular state
of ${\cal H}_i$.  If the state of these degrees of freedom is subsequently
unaltered by further interaction with the system, then different choices
of the $\P_{\alpha_i}$ at time $t_i$ will render the different histories
orthogonal, and hence consistent, at all later times.

This gives us the following picture:  the total Hilbert space for the
system and environment has the form ${\cal H} =
{\cal H}_{\rm sys} \otimes {\cal H}_1 \otimes \cdots {\cal H}_N$,
and the system only interacts with a single subsystem ${\cal H}_i$
at a given time $t_i$.  We choose a set of projections $\{\P_{\alpha_i}^i\}$
onto the state of ${\cal H}_i$ at each time.  Clearly, any history
composed of such projections must decohere.  These subsystems
${\cal H}_i$ are called ``generalized records'' by Gell-Mann and Hartle,
because they record the choice $\P_{\alpha_i}^i$ for all later times;
they are ``generalized'' because the particular set of degrees of
freedom represented by ${\cal H}_i$  might be very complicated, and
inaccessible to an experimenter for all practical purposes
\cite{GMHart3}.

This is highly idealized, of course.  In a real system it is usually
impossible to identify these subsystems ${\cal H}_i$; the choices of
projection are generally not onto these subsystems directly, but
instead depend on correlations in the state of the system and environment;
the interaction with the system does not have this precise,
time-limited form; nor does the system generally interact directly with
the generalized record.  However, in some cases one {\it can} explicitly
identify physical systems which serve as generalized records (for
instance, photons which escape to infinity without interacting further);
and even where one cannot, it is not unreasonable to assume that
something corresponding to them is present \cite{Halliwell}.  It
may be that in a real environment no subsystem remains forever isolated,
but it might be effectively true on
timescales very long compared to the duration of the experiment,
or even the lifetime of the universe.  In particular, situations which
correspond to measurements can almost always be assumed to have some
set of generalized records.  Sometimes these are easy to identify:
lines drawn on graph paper by a measurement apparatus are more than
sufficient to cause decoherence of the measurement results
at all later times.

\section{The model}

\subsection{Model of photodetection}

Photomultiplier tubes are very complicated devices, with the complication
doing little to improve understanding of the detection process.  Therefore
I will use a detector model which is simpler to analyze, based on the
techniques used by Haroche et al.\ \cite{Haroche}
in studying cavity QED in microwave cavities.  Models of this type
have been used before by a number of authors \cite{McElwaine,Kist}.
Figure 1 illustrates the setup.

The heart of our detector is a lossy cavity which is probed by a beam of
atoms.  We assume that the atoms are fairly well localized, so that
their motion can be treated as essentially classical.  One mode of this
cavity can be excited by photons leaking in from the outside; this mode
has creation and annihilation operators $\bdag$ and $\b$.  We assume
that the electronic states of the atoms are prepared in a superposition
of two levels, which we label $\ket0$ and $\ket1$.   If there is an atom
in state $\ket1$ in the cavity it shifts the resonance frequency.  Cavity
loss results from interaction between the mode and the internal degrees of
freedom of the cavity, which we model as a reservoir of harmonic
oscillators.  The time-dependent Hamiltonian for this detector model is
\begin{equation}
\H_{\rm det}(t) = \omega_{\rm det} \bdag\b
  + \sum_j {\rm w}(t-t_j) \bdag\b \ket{1_j} \bra{1_j}
  + \sum_k \gamma_k (\b + \bdag) (\r_k + \rdag_k) + \omega_k \rdag_k \r_k,
\label{detector_H}
\end{equation}
where $\omega_{\rm det}$ is the frequency of the cavity mode,
$\ket{1_j}\bra{1_j}$ is a projector onto state $\ket1$ of the $j$th
atom, $\rdag_k$ and $\r_k$ are the creation and annihilation operators
for the $k$th reservoir oscillator, $\omega_k$ is its frequency, and
$\gamma_k$ is the coupling to the cavity mode.  The function ${\rm w}(t)$
is a window function peaked around $0$, which represents the strength of
the interaction when the atom is inside the cavity.
It is assumed that the traversal
times $t_j$ of the atoms are spaced an average time $\tau$ apart, and
that the width of the window function ${\rm w}(t)$ is narrow compared to
$\tau$.  The important quantity is the integral
\begin{equation}
2 \lambda \equiv \int_{-\infty}^\infty {\rm w}(t)\ dt.
\end{equation}

The effects of the reservoir coupling have been widely studied
and are very well understood.  If we assume that the temperature of
the reservoir is low compared to the oscillator energy $\omega_{\rm det}$
and the frequencies $\omega_k$ are distributed appropriately, the
reservoir produces an effective dissipation rate $\Gamma_1$
\cite{FeynVern,CaldLegg}.

The incoming atoms are prepared in the state
$\ket{\psi_{\rm in}} = (\ket0 + \ket1)/\sqrt2$.  If the cavity mode
is in the number state $\ket{n}$ the atoms leave the cavity in the
state $(\ket0 + \exp(-2i\lambda n)\ket1)/\sqrt2$.  This entanglement
between the atoms and the cavity mode suppresses interference terms
between number states $n$ and $n'$
by a factor of $\cos(\lambda(n-n'))$ per atom, where we assume
$|\lambda(n-n')| < \pi/2$.  With the atoms spaced approximately
$\tau$ apart on average, this corresponds to a decoherence rate
of $\Gamma_2 = -(1/\tau)\ln\cos(\lambda(n-n'))$.

We assume that the dissipation time $1/\Gamma_1$ is long compared to the
decoherence time $1/\Gamma_2$, so that every photon that enters
the cavity persists long enough to be detected, but is short compared to
the rate at which photons enter the cavity, so that there is never
more than one photon in the cavity at a time, and there is essentially
no chance of a photon being coherently reabsorbed by the system from
the detector.  This last point is important for consistency.
In that limit, we can trace out the atomic and reservoir degrees of
freedom, and get an equation for the cavity mode alone,
\begin{eqnarray}
{\dot\rho}_{\rm det} && = - i\omega_{\rm det} [\bdag\b,\rho_{\rm det}]
  + \Gamma_1 \b \rho_{\rm det} \bdag
  - {\Gamma_1\over2} \left(\bdag\b\rho_{\rm det}
  + \rho_{\rm det}\bdag\b \right) \nonumber\\
&& + \Gamma_2 (\bdag\b) \rho_{\rm det} (\bdag\b)
  - {\Gamma_2\over2} \left( (\bdag\b)^2 \rho_{\rm det}
  + \rho_{\rm det} (\bdag\b)^2 \right).
\label{det_master}
\end{eqnarray}
This equation is valid on timescales long compared to $\tau$.

In actual experiments the atoms are themselves measured as they
leave the cavity, but
this is unimportant for our present purpose.  We can imagine them flying
out into space forever, where they form (in the terminology of Gell-Mann
and Hartle) a generalized record of the detector.  Indeed, as we shall
see in section V, the presence of the detector itself is not vital
to derive a quantum trajectory description using decoherent histories.

\subsection{System and output mode}

Consider a quantum system with Hilbert space ${\cal H}_1$, which is
isolated except for a single channel of decay---an interaction
with an external ``output mode,'' which is continuously monitored.
${\cal H}_2$ is the
Hilbert space of the output mode---the cavity mode of (\ref{det_master}).
The combined state of the system plus output mode lies in the product
Hilbert space ${\cal H}_1 \otimes {\cal H}_2$.
This reduced model is illustrated in Figure 2.

The measuring device produces two important
effects.  The first is dissipation.  Excitations of
the output mode will be absorbed by the measuring device at a rate
$\Gamma_1$ which we assume to be rapid compared to the interaction rate
between the system and output mode.
$1/\Gamma_1$ limits the time-resolution of the detector.
The second effect is decoherence.  As the state of the output 
mode becomes correlated with the atom
degrees of freedom in the detector, the phase coherence between
the ground and excited states of the output mode is lost.
This loss of coherence is far quicker than the actual rate of energy loss.
The decoherence rate is $\Gamma_2 \gg \Gamma_1$.

Assume a linear interaction between the system and the output mode,
and go to an interaction picture in the rotating wave approximation.
This lets us remove the Hamiltonian of the output mode, and gives
an interaction Hamiltonian
\begin{equation}
\H_I = \kappa ( \adag \otimes \b + \a \otimes \bdag ).
\end{equation}
The total Hamiltonian is
\begin{equation}
\H = \H_0 \otimes \ident
  + \kappa ( \adag \otimes \b + \a \otimes \bdag ),
\label{Hamiltonian}
\end{equation}
where $\a$ and $\b$ ($\adag$ and $\bdag$) are the lowering (raising)
operators for ${\cal H}_1$ and ${\cal H}_2$, respectively.  We want
dissipation and decoherence to be rapid compared to
the transfer rate between system and output mode.  This
requires that the system should not be too highly excited.
The hierarchy of evolution rates is $\Gamma_2 \gg \Gamma_1
\gg \kappa \expect{\adag\a}$.  We thus have a system with separated
timescales.

The system plus output mode obeys a Markovian master equation:
\begin{eqnarray}
{\dot\rho} &=& - i [\H,\rho] + \Gamma_1 \b \rho \bdag
  - {\Gamma_1\over2} (\bdag\b\rho
  + \rho\bdag\b) \nonumber\\
&& + \Gamma_2 \n_2 \rho \n_2 - {\Gamma_2\over2} (\n_2^2 \rho
  + \rho \n_2^2) = {\cal L} \rho,
\label{total_master}
\end{eqnarray}
where $\rho$ is the density matrix for the combined system and output
mode, and ${\cal L}$ is the Liouville superoperator.
The operator $\n_2 = \bdag\b$ acts on the output mode.
Equation (\ref{total_master}) is linear, and so can be formally solved:
\begin{equation}
\rho(t_2) = \exp\biggl\{ {\cal L}(t_2 - t_1) \biggr\} \rho(t_1).
\end{equation}

Since the output mode is heavily damped, we simplify the problem
by retaining only the two lowest states $\ket0$ and $\ket1$,
treating ${\cal H}_2$ as a two-level system.  We then
expand the density matrix $\rho$ explicitly in terms of its components
in ${\cal H}_1$ and ${\cal H}_2$:
\begin{equation}
\rho(t) = \rho_{00}(t) \otimes \ket0\bra0 + \rho_{01}(t) \otimes \ket0\bra1
  + \rho_{10}(t) \otimes \ket1\bra0 + \rho_{11}(t) \otimes \ket1\bra1,
\end{equation}
where the $\rho_{ij}$ are operators on ${\cal H}_1$ and the $\ket{i}\bra{j}$
on ${\cal H}_2$.  In terms of these components the master equation becomes
\begin{eqnarray}
{\dot\rho}_{00} = && - i[\H_0,\rho_{00}] - i \kappa \adag \rho_{10}
  + i \kappa \rho_{01} \a + \Gamma_1 \rho_{11}, \nonumber\\
{\dot\rho}_{01} = && - i[\H_0,\rho_{01}] - i \kappa \adag \rho_{11}
  + i \kappa \rho_{00} \adag - {\Gamma_1+\Gamma_2\over2} \rho_{01}
  = {\dot\rho}_{10}^\dagger, \nonumber\\
{\dot\rho}_{11} = && - i[\H_0,\rho_{11}] - i \kappa \a \rho_{01}
  + i \kappa \rho_{10} \adag - \Gamma_1 \rho_{11},
\end{eqnarray}

This model may seem highly simplified compared to an actual
photodetection experiment, but it captures most of the essential
physical principles without bogging down in unnecessary detail.  To
discuss photoemission, in is necessary to include some of the environment
degrees of freedom explicitly.  This is the function served by the
output mode, which is about as simple as a physical system can be.
The other important feature is the separation of
timescales between the measuring device degrees of
freedom and the effective dynamical timescale of the system.
The usual approximation of ideal measurements as instantaneous depends
on this separation of timescales.
This sort of model has been used in the past to study the spectrum
of emissions and time-correlation functions of an open optical system
\cite{Schack,Brun2}.

The important element in analyzing this model is its time evolution.  Given
that $\Gamma_1 \ll \Gamma_2$,
it is convenient to expand the time-evolution superoperator
in the following form:
\begin{eqnarray}
&& \e^{\L\delta t} = \e^{\L_2 \delta t}
  + \int_0^{\delta t} dt' \e^{\L_2(\delta t - t')} \L_1 \e^{\L_2 t'} \nonumber\\
&& + \int_0^{\delta t} dt' \int_{t'}^{\delta t} dt'' \e^{\L_2(\delta t - t'')}
  \L_1 \e^{\L_2(t'' - t')} \L_1 \e^{\L_2 t'} + \cdots,
\label{expansion}
\end{eqnarray}
where multiplication of superoperators is composition, with the earliest
rightmost.  Second order terms are all that will be needed in this paper.
Here the terms of the master equation have
been separated:
\begin{equation}
\L = \L_1 + \L_2,
\end{equation}
with
\begin{equation}
\L_1\rho = - i [\H,\rho] + \Gamma_1 \b\rho\bdag
  - {\Gamma_1\over2}(\bdag\b\rho + \rho\bdag\b),
\end{equation}
and
\begin{equation}
\L_2\rho = \Gamma_2 \n_2 \rho \n_2
  - {\Gamma_2\over2} ( \n_2^2 \rho + \rho \n_2^2 ).
\end{equation}
The effects of the superoperators $\L_1$ and $\e^{\L_2 t}$ are
given by
\begin{eqnarray}
(\L_1\rho)_{00} && = - i[\H_0,\rho_{00}] - i\kappa\adag\rho_{10}
  + i\kappa\rho_{01}\a + \Gamma_1 \rho_{11}, \nonumber\\
(\L_1\rho)_{01} && = - i[\H_0,\rho_{01}] - i\kappa\adag\rho_{11}
  + i\kappa\rho_{00}\adag - {\Gamma_1\over2} \rho_{01}
  = (\L_1\rho)_{10}^\dagger, \nonumber\\
(\L_1\rho)_{11} && = - i[\H_0,\rho_{11}] - i\kappa\a\rho_{01}
  + i\kappa\rho_{10}\adag - \Gamma_1 \rho_{11}, \nonumber\\
(\e^{\L_2 t}\rho)_{00} && = \rho_{00}, \nonumber\\
(\e^{\L_2 t}\rho)_{01} && = (\e^{- \Gamma_2 t/2})\rho_{01} =
  (\e^{\L_2 t}\rho)_{10}^\dagger, \nonumber\\
(\e^{\L_2 t}\rho)_{11} && = \rho_{11}.
\label{evolutions}
\end{eqnarray}

Since the superoperator $\e^{\L_2 t}$ is diagonal in the components of
$\rho$, it is particularly simple to insert these expressions into the
expansion (\ref{expansion}) and carry out the integrals explicitly.
If we examine the evolution after a time $\delta t$ where $\Gamma_1 \delta t
\ll 1 \ll \Gamma_2 \delta t$, we see that the off-diagonal terms
$\rho_{01}, \rho_{10}$ are highly suppressed:

\begin{eqnarray}
(\e^{\L\delta t}\rho)_{00} && \approx \rho_{00} - i[\H_0,\rho_{00}]\delta t
  + \Gamma_1 \rho_{11} \delta t
  + {2\kappa^2\delta t\over\Gamma_2}(2\adag\rho_{11}\a
  - \adag\a\rho_{00} - \rho_{00}\adag\a) \nonumber\\
&& + {2\over\Gamma_2}(i\kappa\rho_{01}\a
  - i\kappa\adag\rho_{10} ) - {2i\delta t\over\Gamma_2}[\H_0,i\kappa\rho_{01}\a
  - i\kappa\adag\rho_{10} ]
  - {2\Gamma_1\delta t\over\Gamma_2}(i\kappa\a\rho_{01}
  - i\kappa\rho_{10}\adag) \nonumber\\
&& - [\H_0,[\H_0,\rho_{00}]]\delta t^2/2 - \Gamma_1\rho_{11}\delta t^2/2
  - i [\H_0,\Gamma_1\rho_{11}]\delta t^2 + O(\kappa^2/\Gamma_2^2),
\label{diag0}
\end{eqnarray}

\begin{equation}
(\e^{\L\delta t}\rho)_{01} \approx {2\over\Gamma_2} (i\kappa\rho_{00}\adag
  - i\kappa\adag\rho_{11}) - {2i\kappa\delta t\over\Gamma_2}
  (- i \adag[\H_0,\rho_{11}] + i [\H_0,\rho_{00}]\adag
  - \Gamma_1 [\adag,\rho_{11}] ),
\label{off_diag}
\end{equation}

\begin{eqnarray}
(\e^{\L\delta t}\rho)_{11} && \approx \rho_{11} - i[\H_0,\rho_{11}]\delta t
  - \Gamma_1 \rho_{11} \delta t
  + {2\kappa^2\delta t\over\Gamma_2}(2\a\rho_{00}\adag
  - \a\adag\rho_{11} - \rho_{11}\a\adag) \nonumber\\
&& + {2\over\Gamma_2}(i\kappa\rho_{10}\adag
  - i\kappa\a\rho_{01} ) - {2i\delta t\over\Gamma_2}[\H_0,i\kappa\rho_{10}\adag
  - i\kappa\a\rho_{01} ]
  + {2\Gamma_1\delta t\over\Gamma_2}(i\kappa\a\rho_{01}
  - i\kappa\rho_{10}\adag) \nonumber\\
&& - [\H_0,[\H_0,\rho_{11}]]\delta t^2/2 + \Gamma_1\rho_{11}\delta t^2/2
  + i [\H_0,\Gamma_1\rho_{11}]\delta t^2 + O(\kappa^2/\Gamma_2^2).
\label{diag1}
\end{eqnarray}

The off-diagonal terms $\rho_{01},\rho_{10}$ will always be of order
$O(\kappa/\Gamma_2)$.  We can therefore consider an approximate set of
differential equations in terms of $\rho_{00}$ and $\rho_{11}$ alone:
\begin{eqnarray}
{\dot\rho_{00}} && = - i \Heff \rho_{00} + i \rho_{00} \Heff^\dagger
  + \gamma \adag\rho_{11}\a + \Gamma_1 \rho_{11}, \nonumber\\
{\dot\rho_{11}} && = - i \Heff \rho_{11} + i \rho_{11} \Heff^\dagger
  + \gamma \a\rho_{00}\adag
  - (\Gamma_1 + \gamma) \rho_{11},
\label{intermediate}
\end{eqnarray}
where $\gamma = 4\kappa^2/\Gamma_2$ and the effective Hamiltonian
\begin{equation}
\Heff = \H_0 - i {\gamma\over2} \adag\a
\end{equation}
is the same as that which appears in the quantum jumps formalism.
This equation (\ref{intermediate}) is valid on timescales long compared
to $1/\Gamma_2$.  We examine it further in section IIID below.

If $\Gamma_1$ is large compared to the other terms of the equation, then
we can make the same sort of argument to show that $\rho_{11}$ will
be highly suppressed (by a factor of roughly $\gamma/\Gamma_1$) compared
to $\rho_{00}$.
In this limit we can therefore adiabatically eliminate
all components other than $\rho_{00}$ \cite{Wiseman2}.  If we then
consider a reduced density matrix $\rhot$
for the system alone, without the
output mode, it will be essentially equal to $\rho_{00}$
(with a correction of order $\gamma/\Gamma_1$ from $\rho_{11}$).
The equation for $\rhot$ then becomes
\begin{equation}
{\dot\rhot} = - i [\H_0,\rhot]
  + \gamma \a \rhot \adag
  - (\gamma/2) \adag\a \rhot
  - (\gamma/2) \rhot \adag\a,
\label{adiabatic}
\end{equation}
to first order in $\gamma$.
This equation holds good on time scales long
compared to $\Gamma_1$.
Thus, in the adiabatic limit we see that this indirect measurement
scheme for the total system and output mode
does reproduce the usual master equation
(\ref{cavity}) for the system alone.  Because
$\gamma$ is small, the damping is weak.
This weakness is related to the quantum Zeno effect:  if the
detector had infinite time resolution, the system would
never emit a photon at all \cite{Misra}.

\subsection{Quantum jumps and continuous measurements}

We can unravel equation (\ref{adiabatic}) as described in section
IIB into stochastic trajectories $\ket{{\tilde \psi}(t,\xi)}$.
By averaging $\ket{\tilde\psi}\bra{\tilde\psi}$ over all realizations of
$\xi(t)$ with an appropriate probability measure (\ref{jumps_prob}),
one sees that it does reproduce the master equation
(\ref{adiabatic}) as required \cite{Gardiner}.

The master equation (\ref{adiabatic}) is valid only as long as
the Markovian approximation remains good.  In the case of our toy model,
this means that it is valid only on timescales longer than
$1/\Gamma_1$.  Thus, rather than a jump occurring at a time
$t_i$, it is more correct to consider the jump as occurring during an interval
$\Delta t \sim 1/\Gamma_1$ centered on $t_i$.  For comparison
to experiment this qualification is unimportant,
but it will prove important in making
comparisons to consistent histories.

In the context of photon-counting experiments one can give a
simple physical interpretation to the individual quantum jump trajectories,
as the state of the system conditioned on the continuous measurement
record from the photon counter.  As time passes without the
detection of a photon we gain information about the state of the system;
the lower states become more probable relative to the higher states,
this effect given by the non-Hermitian part of the effective
Hamiltonian.  The jumps represent actual photon detections, in which both
the state of the system and the state of our knowledge change abruptly.

Consider the system plus output mode with Hamiltonian (\ref{Hamiltonian}),
and suppose that von Neumann measurements of the observable $\n_2$
are performed repeatedly on the output mode, separated by time intervals
$1/\Gamma_2$.  If we average over the two possible measurement outcomes,
then the resulting mixed state is
\begin{equation}
\rho \rightarrow \P_0 \rho \P_0 + \P_1 \rho \P_1,
\label{project}
\end{equation}
where $\P_0$ and $\P_1$ are projections onto the states $\ket0$, $\ket1$
of the output mode.  The effect of this repeated measurement is to rapidly
suppress the off-diagonal terms of the density operator $\rho$.

The $\Gamma_2$ terms in the master equation (\ref{total_master}) have
the same effect as (\ref{project}) on time scales long compared to
$1/\Gamma_2$, both giving rise to (\ref{intermediate}) in this limit.
One can thus think of these terms being a continuous
approximation to a series of repeated ideal measurements.  However,
terms of this form also arise generically in the study of systems interacting
with environments \cite{Zurek}.
They are not unique to measurements.  Indeed, this
sort of description can be considered as a model of the measurement
process itself:  instead of taking place instantaneously, the measurement
occurs over a short period of time, as the measured system becomes
entangled with the many degrees of freedom of the measuring device.

One must include as well the $\Gamma_1$ terms which cause the emitted
photons to be absorbed (either by the environment or by the measuring
device, depending on the situation described).  This could be modeled
simply by having the repeated measurements reset the output mode to $\ket0$
after measuring it, with a rate of $\Gamma_1$ (i.e., after every $n$th
measurement where $n=\Gamma_2/\Gamma_1$).  The exact value of $\Gamma_1$
is not very important.  It represents the time-resolution of the
measurement device, and all that is required is that photons not be
emitted more rapidly than they can be absorbed.  Indeed, if one merely
assumes that the measurement device resets the state to $\ket0$ immediately
after each measurement, but that the record cannot resolve individual
clicks within an interval smaller than $1/\Gamma_1$, the probability of
a given measurement record will be exactly given by (\ref{jumps_prob});
the evolution of the system conditioned on the measurement record follows
the quantum jumps formalism given in section IIB.

\subsection{Ortho-jumps}

Consider the coupled equations (\ref{intermediate}), which
arise by averaging the full master equation
(\ref{total_master}) over times long compared to $1/\Gamma_2$.
These equations are equivalent to a Lindblad master equation for the
system plus output mode
\begin{equation}
{\dot \rho} = - i [\H_0,\rho] + \sum_{m=1}^3 \Lhat_m \rho \Lhat_m^\dagger
  - {1\over2} \Lhat_m^\dagger \Lhat_m \rho
  - {1\over2} \rho \Lhat_m^\dagger \Lhat_m,
\label{intermediate_master}
\end{equation}
with Lindblad operators
\begin{eqnarray}
\Lhat_1 = \sqrt{\gamma}\, \a \otimes \bdag, \nonumber\\
\Lhat_2 = \sqrt{\gamma}\, \adag \otimes \b, \nonumber\\
\Lhat_3 = \sqrt{\Gamma_1}\, \ident \otimes \b.
\end{eqnarray}
Any density operator of the form $\rho = \rho_{00} \ket0\bra0
+ \rho_{11} \ket1\bra1$ will remain of that form for all time.

We can now unravel this master equation into the orthogonal jump
trajectories of Di\'osi.  For the moment, let us treat the deterministic
and jump terms separately.  Let us also continue to neglect the higher
excitations of the output mode, so that it can be treated simply as
a two-level system.  If the system begins in an initial state
$\ket\psi = \ket\phi \otimes \ket0$, then the ortho-jump equation
(\ref{ortho_jumps}) becomes
\begin{eqnarray}
\ket{d\psi} = \left( - i \H_0 \ket\phi 
  - {\gamma\over2} \adag\a \ket\phi 
  + {\gamma\over2} \expect{\adag\a} \ket\phi \right) \otimes \ket0 dt
  + {\rm jumps}.
\label{ortho_eqn1}
\end{eqnarray}
If the system begins in the state $\ket\psi = \ket\phi \otimes \ket1$
then the equation is nearly identical:
\begin{eqnarray}
\ket{d\psi} = \left( - i \H_0 \ket\phi
  - {\gamma\over2} \adag\a \ket\phi
  + {\gamma\over2} \expect{\adag\a} \ket\phi \right) \otimes \ket1 dt
  + {\rm jumps}.
\label{ortho_eqn2}
\end{eqnarray}
In {\it both cases} we see that the deterministic part of the evolution
is the same on the system part of the state, leaving the output mode
unchanged, and is equivalent to the deterministic part of the
nonlinear quantum jump equation (\ref{nonlinear_jumps}).  What then
are the effects of the jumps?

A system in a state $\ket\psi$ can jump into any of a set of orthogonal
states which form a basis for the subspace spanned by the states
$\ket{\psi_j} = (\Lhat_j - \expect{\Lhat_j})\ket\psi$.  If the system
is initially in the state $\ket\phi \otimes \ket0$ then only $\Lhat_1$
makes a non-vanishing contribution to that space.  Thus, there is only
one type of jump from that initial condition,
\begin{equation}
\ket\phi \otimes \ket0 \rightarrow
  \a\ket\phi \otimes \ket1/\expect{\adag \a}.
\label{up_jump}
\end{equation}
These jumps occur with a mean rate
$M_{\ket\psi}(dN_1) = \gamma\expect{\adag\a} dt$.
Thus, the effect of this
jump on the system is the same as a standard quantum jump, and they
occur with the same rate.

For a state $\ket\phi \otimes \ket1$ the other two Lindblad operators
$\Lhat_2$ and $\Lhat_3$ can give rise to jumps.  However, because
$\Gamma_1 \gg \gamma$, we can to an excellent approximation neglect the
jumps due to $\Lhat_2$, which occur very rarely compared to jumps due
to $\Lhat_3$.  Making this approximation, the jumps take the form
\begin{equation}
\ket\phi \otimes \ket1 \rightarrow \ket\phi \otimes \ket0,
\label{down_jump}
\end{equation}
and occur with a mean rate $M(dN_3) = \Gamma_1$ independent of the state
$\phi$.  These downward jumps happen much more rapidly than the upward
jumps due to $\Lhat_1$, and a trajectory has a near-unity probability
of dropping back down to $\ket0$ within a time of order $1/\Gamma_1$
after jumping up to $\ket1$.

We can thus characterize the typical evolution of these trajectories.
They spend most of their time (by a ratio of roughly $\Gamma_1/\gamma$)
in the lower state $\ket0$, with the system evolving according to the
deterministic terms of equation $(\ref{nonlinear_jumps})$.  At random
intervals, there is a jump in which the system is multiplied by $\a$
and the output mode jumps to $\ket1$.  Within a random time of order
$1/\Gamma_1$ the output mode drops back down to $\ket0$, while the system
continues to evolve according to $(\ref{nonlinear_jumps})$ undisturbed.

If we trace out the output mode, we see that one of these ortho-jump
trajectories is exactly equivalent to a standard jump trajectory on
the system Hilbert space alone.  Each standard jump trajectory will
have many ortho-jump trajectories equivalent to it, each corresponding
to a slightly different time for the output mode to return to the state
$\ket0$.  However, as the exact time
has no effect on the system evolution, this multiplicity is
unimportant.  The total probability of the ortho-jump trajectories
will be exactly the same as the probability of the equivalent standard
jump trajectory.

This many-to-one relationship is a simple example of a {\it coarse-graining};
the ortho-jump trajectories are more detailed than the standard jump
trajectories, because they include information about the state of the
output mode as well.  This concept is very important in the decoherent
histories formulation of the same problem.  What is more, it has already
been shown by Paz and Zurek and Di\'osi \cite{Diosi1,PazZurek}
that orthogonal jump trajectories
are equivalent to a particular set of decoherent histories.  This new
equivalence to standard jumps (true at least in this simple model) makes
it easy to show how standard jumps can be represented by decoherent
histories.

Strictly speaking, the equivalence shown above in equations
(\ref{ortho_eqn1}--\ref{ortho_eqn2}) is for the nonlinear version
of both unravelings.  As mentioned before, in general there is
no linear version of ortho-jumps.  However, when neglecting jumps due
to coherent reabsorption we see that this system is a special case,
with all jumps between the orthogonal subspaces labeled by $\ket0$
and $\ket1$.

Because of this there is a linear version of ortho-jumps for this case,
produced simply by dropping all the nonlinear terms in equations
(\ref{ortho_eqn1}--\ref{ortho_eqn2}) and removing the renormalization
in the jump (\ref{up_jump}).  The resulting equations are identical in
form to the linear version of standard jumps, but the jumps are no
longer strictly identified with ``clicks'' of a detector.  The
coarse-grained version of this linear equation is equivalent to
standard linear quantum jumps on the system alone, just as in the
nonlinear case.

\section{Consistent histories description}

As described in section IID, a particular history
is given by choosing one projection $\P^j_{\alpha_j}(t_j)$ at each
time $t_j$, specified by the sequence of indices $\{\alpha_j\}$ denoted
$h$ for short.  The decoherence functional on a pair of histories
$h$ and $h'$ is then given by
\begin{equation}
D[h,h'] = \Tr \biggl\{ \P^N_{\alpha_N}(t_N) \cdots
  \P^1_{\alpha_1}(t_1) \rho(t_0) \P^1_{\alpha_1'}(t_1) \cdots
  \P^N_{\alpha_N'}(t_N) \biggr\},
\end{equation}
where $\rho(t_0)$ is the initial density matrix of the system
\cite{GMHart1}.

We specialize to the system plus output mode described in section III.
They are initially in the pure state
$\ket\Psi = \ket{\psi_0} \otimes \ket0$.  Since
the degrees of freedom of the environment (e.g., the internal degrees of
freedom of the measuring device) have already been traced out, we
replace the simple Schr\"odinger evolution (\ref{schrodinger}) with the
Liouvillian evolution of master equation (\ref{total_master}),
according to the quantum regression theorem \cite{QRT}.
The decoherence functional for two histories $h$ and $h'$ then has the form
\begin{equation}
D[h,h'] = \Tr \biggl\{ \P_{\alpha_N} \e^{\L\delta t}(
  \P_{\alpha_{N-1}} \e^{\L\delta t}( \cdots \e^{\L\delta t}(
  \P_{\alpha_1} \ket\Psi\bra\Psi \P_{\alpha_1'} )
  \cdots ) \P_{\alpha'_{N-1}} ) \P_{\alpha_N'} \biggr\},
\label{jump_functional}
\end{equation}
where $\L$ is the Liouville superoperator from (\ref{total_master}).

We consider histories composed
of the following Schr\"odinger projections:
\begin{equation}
\P_0 = \ident \otimes \ket0\bra0,\ \ 
  \P_1 = \ident \otimes \ket1\bra1.
\end{equation}
These projections represent the absence or presence of a photon in the
output mode.  These projections are spaced a short time $\delta t$
apart, and each history is composed of $N$ projections, representing a
total time $T = N\delta t$.  A single history $h$ is given by the string
$\{\alpha_1,\alpha_2,\ldots,\alpha_N\}$, where $\alpha_j = 0,1$ represents
whether or not a photon has been emitted at time $t_j = (j-1) \delta t$.
The time-evolution superoperators in (\ref{jump_functional})
tend to evolve pure states
to mixed states.  This is counteracted by the effect of the
repeated projections $\P_\alpha$, as we shall see.  There are two
important issues to address within the consistent histories formalism:
the probabilities of histories (given by the diagonal terms of the
decoherence functional) and the decoherence of the set of histories as
a whole (given by the off-diagonal terms).  We examine them separately.

\subsection{Probability of histories}

From the expressions (\ref{expansion}--\ref{evolutions}),
we can determine the character of the
different histories.  The crucial choice is the size
of the spacing $\delta t$ between projections.
Too small and the histories will
not decohere.  Too large and all we will see will be standard master
equation evolution, unresolved into trajectories.  The interesting regime
is in the range
\begin{equation}
{1\over\Gamma_2} \ll \delta t \ll {1\over\Gamma_1}
\end{equation}
as described in equations (\ref{diag0}--\ref{intermediate}).
On this timescale, the $\Gamma_2$ terms are sufficient to insure decoherence
while the effects of the $\Gamma_1$
terms are resolved into individual trajectories.  The time-evolution
produced by the $\exp(\L\delta t)$ superoperators is given by the
simple equations (\ref{intermediate}), equivalent to the averaged
Lindblad equation (\ref{intermediate_master}).

If the external mode is initially unexcited, with
$\rho = \rho_{00} \otimes \ket0 \bra0$, then
after evolving for a time $\delta t$ the state becomes
\begin{eqnarray}
(\e^{\L\delta t}\rho)_{00} && 
  \approx \e^{ - i (\H_0 - i\gamma\adag\a/2 ) \delta t} \rho_{00}
  \e^{ i (\H_0 + i\gamma\adag\a/2 ) \delta t}, \nonumber\\
(\e^{\L\delta t}\rho)_{01} && \approx O(\kappa/\Gamma_2), \nonumber\\
(\e^{\L\delta t}\rho)_{11} && \approx \gamma \a \rho_{00} \adag \delta t.
\label{unexcited}
\end{eqnarray}
Here we see the appearance of the effective Hamiltonian $\Heff$, just
as in the quantum jump unraveling.

We can also consider the case when the external mode is initially excited,
with $\rho = \rho_{11} \otimes \ket1 \bra1$.  After a time $\delta t$
the state becomes
\begin{eqnarray}
(\e^{\L\delta t}\rho)_{00} && \approx \Gamma_1 \rho_{11} \delta t
  + \gamma \adag \rho_{11} \a \delta t, \nonumber\\
(\e^{\L\delta t}\rho)_{01} && \approx O(\kappa/\Gamma_2), \nonumber\\
(\e^{\L\delta t}\rho)_{11} &&
  \approx (1 - (\Gamma_1 + \gamma)\delta t)
  \e^{ - i \Heff \delta t} \rho_{11}
  \e^{ i \Heff^\dagger \delta t}.
\label{excited}
\end{eqnarray}
Once again the effective Hamiltonian appears, together with two additional
effects.  The first is the possibility that the photon in the excited
mode will be absorbed by the measuring device.  The second (much smaller)
effect is the possibility that the photon will be coherently re-absorbed
by the system.  This process is so weak as to be negligible within the
regime we are considering, and we will henceforth neglect it.
This is the same approximation made in the ortho-jumps unraveling
(\ref{intermediate_master}--\ref{down_jump}) by neglecting the
$\Lhat_2$ contribution to the jumps.

By combining the above expressions with the appropriate
projections $\P_0$ and $\P_1$ (which pick out the $\rho_{00}$ or
$\rho_{11}$ component, respectively), we can write down the probabilities
of different possible histories.
Let us examine three illustrative
cases and see how they exactly parallel quantum jump trajectories.

\subsubsection{Evolution without jumps}

Suppose that initially $\rho_{00} = \ket\psi \bra\psi$ while
$\rho_{01} = \rho_{10} = \rho_{11} = 0$, i.e., the system is in a pure
state and no photon has been emitted.  Let us consider the history
given by an unbroken string of $N$ $\P_0$ projections, corresponding
to no photon being emitted during a time $N\delta t$.

The probability of such a history is given by the diagonal element
$D[0^N,0^N]$ of (\ref{jump_functional}).  We can pick out the $\rho_{00}$
component of (\ref{unexcited}), and see that
after the first time interval $\delta t$
\begin{equation}
\P_0 \e^{\L\delta t} (\ket{\psi}\bra{\psi} \otimes \ket0 \bra0) \P_0
  \approx
\biggl(\e^{-i \Heff \delta t} \ket\psi
  \bra\psi \e^{i \Heff^\dagger \adag\a) \delta t} \biggr)
  \otimes \ket0 \bra0.
\end{equation}
Repeating this $N$ times and tracing out the output mode
Hilbert space ${\cal H}_2$ we get
\begin{equation}
D[0^N,0^N] \approx \Tr_1 \biggl\{ \e^{- i \Heff N\delta t} \ket\psi
  \bra\psi \e^{i \Heff^\dagger N\delta t} \biggr\},
\label{no_jumps}
\end{equation}
which exactly agrees with the probability of the quantum jump or
ortho-jump trajectories
when no jumps are detected. ($\Tr_1$ indicates a trace over
${\cal H}_1$ only.)

\subsubsection{Evolution up to a single jump at time $N\delta t$}

Here we can make use of the previous result (\ref{no_jumps}) up until
time $N\delta t$, when instead of using projection $\P_0$ we use
$\P_1$.  This is the same as keeping the $\rho_{11}$ component
of $\exp(\L\delta t)\rho$ instead of the $\rho_{00}$ component at the
final projection time.  This yields
\begin{equation}
D[0^N 1,0^N 1] \approx (\gamma\delta t) \Tr_1 \biggl\{
  \a \e^{- i \Heff N\delta t} \ket\psi
  \bra\psi \e^{i \Heff^\dagger N\delta t} \adag \biggr\},
\label{one_jump}
\end{equation}

Once again, this agrees with the probability of the corresponding
quantum jump trajectory.  (Note, though, that in this context, $\delta t$
is short compared to the $\Delta t$ in expression (\ref{jumps_prob}).  This
is due to the finer-grained nature of this history, about which more below.)

\subsubsection{Evolution after a jump}

What happens after the external mode has ``registered'' as being in the
excited state?  Essentially, there are two possibilities:  either the
external mode can drop back down to the unexcited state (representing
absorption of the photon by the measuring device) or it will remain in
the excited state.  We can examine these two possibilities separately:
\begin{equation}
\P_0 \e^{\L\delta t} (\ket{\psi'} \bra{\psi'} \otimes \ket1 \bra1) \P_0
  \approx \Gamma_1 \delta t \ket{\psi'} \bra{\psi'} \otimes \ket0 \bra0,
\label{after_jump0}
\end{equation}
\begin{equation}
\P_1 \e^{\L\delta t} (\ket{\psi'} \bra{\psi'} \otimes \ket1 \bra1) \P_1
  \approx (1 - \Gamma_1 \delta t) \e^{-i\Heff\delta t} \ket{\psi'}
  \bra{\psi'} \e^{i\Heff^\dagger\delta t} \otimes \ket1\bra1.
\label{after_jump1}
\end{equation}

So we see that the external mode has a probability of roughly
$\Gamma_1\delta t$ per time $\delta t$ of dropping back down to the ground
state, whereupon it resumes evolution as in (\ref{no_jumps}), and a
probability of $1-\Gamma_1\delta t$ of remaining in the excited state.
In either case, the system component of the state continues to evolve
according to the effective Hamiltonian $\Heff$.

This is exactly the same situation we encountered with ortho-jumps
in section IIID.  There, we saw that a typical trajectory spent the
majority of time with the output mode in the ground state. Infrequently
this mode absorbs a photon from the system and becomes excited;  this
photon is dissipated into the environment in a time of order $1/\Gamma_1$.
The rate for coherent reabsorption by the system is much lower, and can
be neglected.

Thus, we see that this set of histories corresponds exactly to the set of
ortho-jump trajectories at the intermediate timescale $\delta t$.  This is,
in fact, a special case of the correspondence shown by Di\'osi.  In his
treatment, the histories were {\it branch-dependent} in order to preserve
the orthogonality of the jumps.  In ours, the jumps are always between
two known subspaces, and thus are branch independent.

To reproduce standard jumps, we must coarse-grain just as before.
Consider an interval
$\Delta t = M\delta t$
which is long compared to $1/\Gamma_1$ but still short compared to the
dynamical timescales of the system.  Let $\omega$ be a frequency which
characterizes the system Hamiltonian $\H_0$.  Then we have
$\Gamma_1 \Delta t \gg 1 \gg \omega \Delta t \gg \omega \delta t$.
Let us sum all histories which include one jump and re-absorption within
this interval (the probability of multiple jumps is small enough to be
neglected), and look at the time-propagators occurring inside the
trace:
\begin{eqnarray}
\sum_{j=0}^M  && \Tr_1 \left\{ \cdots \e^{-i\Heff j\delta t} \a
  \e^{-i\Heff(M-j)\delta t} \cdots \ket\psi \bra\psi \cdots
  \e^{i\Heff\d(M-j)\delta t} \adag
  \e^{i\Heff\d j\delta t} \cdots \right\} \nonumber\\
&& \approx (M+1) \Tr_1 \left\{ \cdots \e^{-i\Heff M\delta t/2} \a
  \e^{-i\Heff M\delta t/2} \cdots \ket\psi \bra\psi \cdots
  \e^{i\Heff\d M\delta t/2} \adag
  \e^{i\Heff\d M\delta t/2} \cdots \right\},
\label{coarse_grain}
\end{eqnarray}
with an error of order $M\omega^2\Delta t^2$.  The exact absorption time
is irrelevant, since the external mode is traced over, the system
dynamics are essentially unaffected, and the probabilities sum to 1.  So making
this coarse-graining, and combining the above cases gives an expression for
the probability of a particular jump record of
\begin{eqnarray}
p(t_1,\ldots,t_n) = &&  (\gamma\Delta t)^N
  \Tr\biggl\{ \e^{-i\Heff(T-t_N)} \a \e^{-i\Heff(t_N - t_{N-1})} \a
  \cdots a \e^{-i\Heff t_1} \nonumber\\
&& \times \ket\psi\bra\psi \e^{i\Heff^\dagger t_1} \adag
  \cdots \adag \e^{i\Heff^\dagger(T-t_N)} \biggr\},
\label{history_prob}
\end{eqnarray}
which exactly matches the quantum jump expression
(\ref{jumps_prob}).

\subsection{Decoherence of histories}

Such a histories description is only meaningful if
the histories decohere.  Exact consistency, as in
(\ref{decoherence}), is a difficult criterion to meet.  It is
more usual to show that a model is {\it approximately} consistent,
which generally insures that the histories satisfy the probability sum
rules to some level of precision.

One criterion for approximate consistency has been suggested by
Dowker and Halliwell \cite{DowkHall}.  If we wish the probability sum
rules to be satisfied to a precision $\epsilon \ll 1$, we require that
\begin{equation}
|D[h,h']|^2 < \epsilon^2 D[h,h] D[h',h'] = \epsilon^2 p(h) p(h'),
\end{equation}
for all unequal pairs of histories $h,h'$.  Generally speaking, the
``more different'' a pair of histories is (i.e., the more projections
they differ in), the more suppressed the off-diagonal term.
This is certainly true for this model of photodetection.  So it
suffices to look at two histories which are as close as possible
without being identical.

In the case of these ``jump'' histories, this means that these histories
differ at a single time $t_i$, one having a projection $\P_0$, the
other $\P_1$.  In the decoherence functional, this is equivalent to
picking out the $\rho_{01}$ or $\rho_{10}$ component of
$\exp(\L\delta t)\ket{\psi'}\bra{\psi'}$ at that time, sandwiched
between identical projectors $\P_0$ or $\P_1$ on either side.

Examining the components given by (\ref{expansion}--\ref{evolutions})
and (\ref{unexcited}--\ref{excited}),
we see that
\begin{equation}
{ |D[h,h']|^2 \over p(h) p(h') } \sim { 1 \over (\Gamma_2 \delta t)^2 },
\end{equation}
so we expect the sum rules to be obeyed with a precision of roughly
$O(1/\Gamma_2\delta t)$.  For large $\Gamma_2$ this is more than
adequate.

\subsection{Partial trace decoherence}

Finkelstein \cite{Finkelstein} has suggested an interesting alternative
definition for decoherence in the case of a system-environment split,
which he terms PT or ``Partial Trace'' decoherence.  Consider the
operator-valued functional defined by
\begin{equation}
\Dbar[h,h'] = \Tr_{\rm env} \biggl\{ \P^N_{\alpha_N}(t_N) \cdots
  \P^1_{\alpha_1}(t_1) \rho(t_0) \P^1_{\alpha_1'}(t_1) \cdots
  \P^N_{\alpha_N'} \biggr\},
\end{equation}
where $\Tr_{\rm env}$ denotes a partial trace over the environment degrees of
freedom only.  $\Dbar[h,h']$ is therefore an operator in the Hilbert
space ${\cal H}_1 \otimes {\cal H}_2$.  The criterion for PT decoherence is
\begin{equation}
\Dbar[h,h'] = 0, h \ne h',
\label{pt_decoherence}
\end{equation}
which is a much stronger condition than the usual decoherence criterion;
it means that the different alternative histories are orthogonal in
the environment degrees of freedom alone, implying the existence of
generalized records \cite{GMHart3}.

From the form of the standard decoherence functional and the correspondence
with the ortho-jumps unraveling (\ref{ortho_eqn1}--\ref{ortho_eqn2}),
we see that $\Dbar[h,h]$ will be a projector onto the state of the system
plus output mode corresponding to the particular trajectory $j$.  And from
the equivalence of this to standard jumps, it follows that
\begin{equation}
\Dbar[h,h] = \ket{\tilde\psi_h} p(h) \bra{\tilde\psi_h}
  \otimes (\ket0\bra0\  {\rm or}\  \ket1\bra1),
\end{equation}
where $\ket{\tilde\psi_h}$ is the normalized state produced by the 
quantum jump trajectory corresponding to the
history $h$, and $p(h)$ is the probability of this trajectory.
By exactly the same argument as before, the off-diagonal terms vanish
and therefore this system is PT decoherent.

\section{Generalized Records}

As discussed in section IIB, one of the most powerful sources of
decoherence is the creation of entanglement between ``system''
and ``environment'' degrees of freedom.  Because these correlations
exist, a careful measurement of the environment could in principle
reveal the exact state of the system; thus, the different possible
system states cannot interfere, and condition (\ref{decoherence})
is assured.  Gell-Mann and Hartle term these persistent correlations
``generalized records;'' they are generalized because in practice
the measurement needed to access them might be impossibly difficult.

For instance, suppose a measuring device records a series of results
onto a paper tape.  This is a record in the usual sense, and its
existence guarantees decoherence between different possible measurement
outcomes.  If we then burn the tape, the record becomes difficult to
access.  But nevertheless, in principle the information is not lost; by
cleverly measuring the ashes, air, and outgoing light one might in
principle be able to reconstruct it, and thus decoherence persists.
This ``in principle'' recoverable information is a generalized record.

Suppose our system and environment begin in an initial pure state.
Then the decoherence functional becomes (\ref{inner_product}),
implying that the unnormalized states $\{\C_h\ket\Psi\}$ are all
orthogonal (or zero).  This implies the existence of a set of
orthogonal projection operators $\R_h$ such that
$\R_h\C_{h'}\ket\Psi = \delta_{hh'} \C_h\ket\Psi$.  Any such collection
of orthogonal projection operators corresponds to an observable.
Thus, in the pure state case, decoherence implies the existence of
an observable whose value is exactly correlated with which history
occurred.  Conversely, the existence of such an observable implies
decoherence.  This converse holds even in the case of mixed states.
This observable is the generalized record, and its existence is a
stronger consistency criterion than (\ref{decoherence}).  If the
record is contained entirely in the environment degrees of freedom,
so $\R_h = \ident_{\rm sys}\otimes\R_h^{\rm env}$
where $\R_h^{\rm env}$ is a projector
on the environment Hilbert space alone, it implies the partial trace
decoherence (\ref{pt_decoherence}) of Finkelstein.

In the case of our model, the outgoing atoms clearly serve as such
a generalized record.  Suppose the output mode is in the ground state
throughout some period $\Delta t$.  In that case, all the outgoing
atoms should be in the state $(\ket0+\ket1)/\sqrt2$.  If the output
mode is instead in the excited state, the probability is overwhelming
that at least some of the atoms will be found in the state
$(\ket0 - \ket1)/\sqrt2$.  Thus, these atoms serve as a record of
the state of the output mode during the period $\Delta t$.

Of course, if the output mode were only in the excited state for a
short period, the atoms might fail to record the fact.  But since
photons in the output mode persist for an average time of order
$1/\Gamma_1 \gg \Delta t$, this ambiguity is unimportant in practice.

What is more, projections on the states of different atoms commute
with each other.  Thus we might, in principle, make no assertions about
the state of the output mode, but merely wait until a large number
of atoms had passed through the detector and then project onto their
collective state.  This projection would be sufficient to reconstruct
the quantum trajectory in its entirety.

In fact, we can play even more elaborate games than that.  There is
no reason, after all, why we should use the same projections on all
the atoms.  We might switch from one set to another as we chose.  In
the language of quantum trajectories, this would be like switching
our choice of unraveling at different times.  Indeed, we could go further
than this:  we could make our choice of projections on some atoms
contingent on our results for other atoms.  Thus, our choice of unraveling
at one time might depend on the state at another time.  What is more,
there is no reason that we could not choose our unraveling at early
times (i.e., the first atoms to pass through) contingent upon results
at later times (i.e., the last atoms to pass through), rather than the
other way around.  Ordinary quantum trajectories theory does not consider
this possibility for good reason, since such a measurement scheme would
be extremely difficult in practice, nor is it obvious what is gained
from it, but we can see how the fairly
specific assumptions of quantum trajectories can be progressively
broadened towards the greater generality of consistent histories.
The existence of generalized records insures that our set of histories
will be consistent under all such permutations.

The existence of generalized records also lets us get away from the
whole notion of measurement.  Suppose that instead of putting a
photodetector outside our cavity we put nothing there, and allow
emitted photons to escape to infinity.  Away from the immediate vicinity
of the system, we can describe this external electromagnetic fields
in terms of incoming and outgoing plane waves with creation and
annihilation operators $\adag_{\pm k\ell}$ and $\a_{\pm k\ell}$,
where $\ell$ labels the polarization.  A reasonable initial condition
would be that the incoming field is in the vacuum state.  The outgoing
field carries away information about the system.  Using plane waves
makes it difficult to discuss properties which are local in time;
consider instead the operators $\c_{x_0k_0\ell}$
and $\cdag_{x_0k_0\ell}$ defined by
\begin{equation}
\c_{x_0k_0\ell} = ({\rm const})\times
  \int dk \exp\left\{ - {(k-k_0)^2\over\Delta k^2}
  + i k x_0 \right\} \a_{k\ell},
\end{equation}
which annihilate (create) a Gaussian wave packet
centered at $x_0$ and $k_0$.  Needless to say, these states
are overcomplete.  But we can still use them to define operators
$\F_C$, which are quasiprojectors onto states with a single photon
in phase space region $C$, and act as the identity
{\it outside} region $C$.  If $C$ has a regular boundary, then as
the volume of $C$ becomes large compared to $\hbar$ $\F_C$ becomes
closer and closer to a true projector.  What is more, if $C$ and
$C'$ are nonoverlapping regions, $F_C$ and $F_{C'}$ approximately
commute \cite{Omnes4}.

Consider now a rectangular region $C$ with length $c\Delta t$ much longer
than the width of the wave packets, located well outside
the near field of the system, and extent in wavenumber space
$l_k \gg \Delta k$ which is large compared to the linewidth of
the emitted photons and limited to outgoing photons, centered on
the cavity wavenumber $\omega_0/c$.  Consider histories with projections
$\F_C$ and $\ident - \F_C$ at a series of times $t_j = j\Delta t$.
The projection $\ident-\F_C$ is essentially a projection on no photon
being emitted, since the probabilities of multiple photons or photons
outside the frequency range are negligibly small.  This set of histories
will correspond on the system space to exactly the set of trajectories
(\ref{history_prob}) we derived before.

Time evolution by $\Delta t$ is just equivalent to shifting the region
$C$ ``outward'' by a distance $c\Delta t$.  Thus the Heisenberg
projections $\F_C(t_j)$ which contribute to the history operator
(\ref{history_op}) are actually projectors onto non-overlapping regions,
and hence commute.  We could replace them all with the single
projector
\begin{equation}
\R_{i_1\ldots i_n} = \F_{i_n}(t_n) \cdots \F_{i_1}(t_1),
\end{equation}
where $i_j = 0,1$ and $\F_1(t_j) = \F_C(t_j)$,
$\F_0(t_j) = \ident - \F_C(t_j)$.  The observable represented by the
set of projections $\R$ is a generalized record of the quantum jumps
trajectory.

However, just as in the case of the photodetector, we can imagine
projecting onto many other things besides photon number, each choice
corresponding to a different unraveling.  As long as we restrict our
projections to outgoing waves, consistency is guaranteed.  If we
instead choose to project onto some combination of incoming and outgoing
waves, the histories would not be consistent, in general.  This is just
as if our measurements of the outgoing field had a chance of reflecting
light back into the system; in that case the condition
(\ref{traj_consistency}) would generally not hold, and the measurement
scheme would no longer be an unraveling of the master equation.

Again, we might choose one set of projections at one time and another
at another, so as to change unravelings, and there is no reason in
principle not to make the choice of projections at one time contingent
on the results at another.

This freedom leads to a somewhat curious situation.  The outgoing
field from the system constitutes a generalized record of the system
evolution.  But what is it a record of?  By choosing different
measurement schemes, or more generally different consistent descriptions,
we can reconstruct very different-looking evolutions.  It has long
been known that on the one hand there were many different, incompatible
unravelings of the master equation, and on the other many different
incompatible sets of consistent histories.  We now see that these can
correspond to different, incompatible generalized records.

\section{Conclusions}

We have seen how a continuous measurement can be described
in terms of decoherent histories, and how the resulting histories
match the corresponding quantum trajectory unraveling of the master 
equation.  The probabilities of the
decoherent histories match the weights of the given trajectories,
and the off-diagonal terms of the decoherence functional are highly
suppressed, as we would expect from a measurement situation.

Decoherent histories, as is widely known, must obey a
condition (\ref{decoherence}) for their probabilities to have
a consistent interpretation.  What has been less widely appreciated
is that quantum trajectories must obey a consistency
condition (\ref{traj_consistency}) of their own.  The reason that
this has received so little remark is that all commonly
considered unravelings of the master equation automatically satisfy
this requirement.  These schemes rely on measurements of an outgoing
field, which constitutes a generalized record of the system evolution,
guaranteeing that both consistency conditions are obeyed.

The existence of multiple incompatible records creates an interesting
ambiguity in our description.  Suppose that instead of measuring the
field as it is emitted from the system, we wait for a long time and
measure the field far away from the system.  Our measurements then
correspond to information about the system evolution long before,
a sort of super-delayed-choice experiment.  The choice of unraveling
might be delayed until long after the evolution itself is complete.
Were there no quantum trajectories until the measurement occurred?
Does one trajectory ``really'' happen?  Do all of them somehow coexist?

These are the sort of interpretational questions
upon which gallons of ink can be spent
without putting an end to argument.  From a practical point of view, it
doesn't matter.  The fact that these sets of histories are consistent
means that we can choose an unraveling and make arguments based on
the trajectories without fear that inconsistencies will invalidate our
results.  While there may be no actual measurement record, in the cases
we consider the existence of generalized records serves the same function.
While these generalized records might be hard to identify, and even
harder to measure (especially if they depart at the speed of light),
they insure that distinct alternatives remain distinct,
and cannot subsequently interfere with each other.
This is a valuable fact, especially since real measurements rarely
approach the level of perfection commonly assumed for pure state
unravelings.  If these unravelings decohere from each other, they can
be treated like classical alternatives without fear of
contradictions arising.  Imperfect measurements can then
be treated by further coarse-graining of these histories.

What is more, if there are many different unravelings corresponding
to different sets of decoherent histories, the choice of which to use
can become a matter of convenience.  In some situations
a jump-like description is most useful; in others, a diffusion equation.
It might be possible to use either one by an appropriate choice of
projection operators on the same environment.

Consistent histories is a powerful formalism, but
producing a full set of histories can be dauntingly difficult, especially
as their number increases exponentially with the number of projections.
In practical calculations, it is often far easier to
randomly choose some subset of ``typical'' histories and average over them.
For this, quantum trajectory equations are supremely well-suited.  Once
again, the freedom to choose among different but equivalent formulations
of a problem is a boon to the theorist.

Over the last several years a great deal of work has been done on problems
of quantum measurement, from a wide variety of viewpoints.  I believe
that these approaches, to the extent that they are correct,
are connected at a deep level, and that by
understanding the connections their power can be enhanced.
This work is a small step in that direction.  I have no doubt that much
more progress is to come.

\section*{Acknowledgments}

I would like to thank Lajos Di\'osi for very valuable
discussions, and Howard Carmichael,
Murray Gell-Mann, Nicolas Gisin, Robert Griffiths,
Jonathan Halliwell, James Hartle, Peter Knight,
Ian Percival, Martin Plenio, and R\"udiger Schack for
suggestions and feedback.  Dieter Zeh asked provocative questions.
Financial support was provided by
the UK EPSRC and by NSF Grants PHY-94-07194 and PHY-96-02084.

\vfil\eject

\vfil

Figure 1.  Picture of the atom-beam photodetector.  The atoms are prepared
in an initial state $\ket\psi = (\ket0 + \ket1)/\sqrt2$ and pass
one at a time through the cavity where they interact briefly with
the field of the cavity mode.  The average time $\tau$ between atoms is
short compared to the lifetime of the cavity mode, which is in turn
short compared to the rate at which photons enter the cavity.  Thus,
on average much less than one photon is present in the cavity.

\vfil

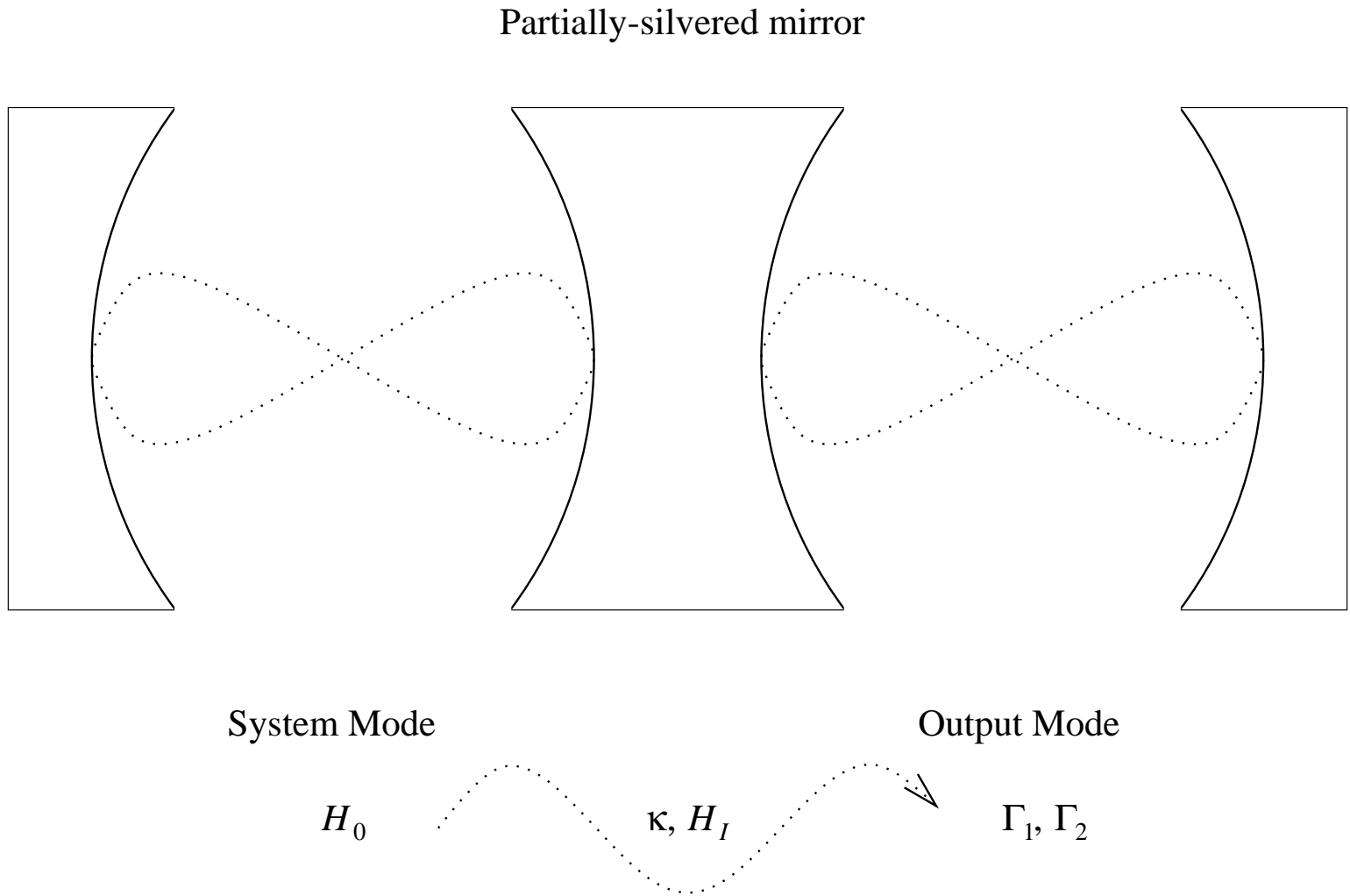
Figure 2.  Schematic picture of the model system.  The system, with
internal Hamiltonian $\H_0$ interacts via an interaction Hamiltonian
$\H_I$ with an output mode, which is modeled as a two-level system.
The interaction is characterized by a rate $\kappa$, and the
output mode is part of a measuring device causing dissipation
and decoherence with rates $\Gamma_1, \Gamma_2$.

\vfil

\end{document}